\documentclass[floatfix,superscriptaddress,twocolumn,showpacs,aps,amsmath,amssymb]{revtex4-1}
\usepackage{graphicx}
\usepackage{amssymb}
\usepackage{color}
\usepackage{bm}
\usepackage{multirow}

\usepackage{times}

\usepackage{xcolor}

\definecolor{darkgreen}{rgb}{0,0.6,0}

\def\tc{$T_\mathrm{C}$\,}

\newcommand\Tsc{T_\mathrm{sc}}
\newcommand\TC{T_\mathrm{C}}

\begin{document}

\title{Piezomagnetism in the Ising ferromagnet URhGe}

\author{Mikiya~Tomikawa}\email{tomikawa.mikiya.76a@kyoto-u.jp}
\affiliation{Department of Physics, Graduate School of Science, Kyoto University, Kyoto 606-8502, Japan}

\author{Ryo~Araki}
\affiliation{Department of Physics, Graduate School of Science, Kyoto University, Kyoto 606-8502, Japan}

\author{Atsutoshi~Ikeda}
\affiliation{Department of Electronic Science and Engineering, Graduate School of Engineering,  Kyoto University, Kyoto 615-8510, Japan}

\author{Ai~Nakamura}
\affiliation{Department of Physics, Graduate School of Science, IMR, Tohoku University, Oarai, Ibaraki 311-1313, Japan}

\author{Dai~Aoki}
\affiliation{Department of Physics, Graduate School of Science, IMR, Tohoku University, Oarai, Ibaraki 311-1313, Japan}
\affiliation{CEA-Grenoble, 38000 Grenoble, France}

\author{Kenji~Ishida}
\affiliation{Department of Physics, Graduate School of Science, Kyoto University, Kyoto 606-8502, Japan}

\author{Shingo~Yonezawa}\email{yonezawa.shingo.3m@kyoto-u.ac.jp}
\affiliation{Department of Physics, Graduate School of Science,  Kyoto University, Kyoto 606-8502, Japan}
\affiliation{Department of Electronic Science and Engineering, Graduate School of Engineering,  Kyoto University, Kyoto 615-8510, Japan}

\date{\today}

\begin{abstract}
Piezomagnetism, linear response between strain and magnetic field, is relatively unexplored cross-correlation but has promising potential as a novel probe of time-reversal-symmetry breaking in various classes of materials.
Interestingly, there has been no report of piezomagnetism in ferromagnets, most archetypal time-reversal-symmetry-broken materials.
This half-century absence of piezomagnetic ferromagnets is attributable to complications originating from multiple-domain states, as well as from changes in the magnetic point group by rotation of magnetic moment.
Here, we report characteristic V-shaped magnetostriction in the Ising itinerant ferromagnet URhGe, observed by simultaneous multi-axis strain measurement technique utilizing optical fiber Bragg grating sensors.
This novel magnetostriction occurs only under fields along the $c$ axis and does not scale with the square of magnetization.
Such unconventional feature indicates piezomagnetism as its origin.
Our observation, marking the first report of piezomagnetism in ferromagnets, is owing to the mono-domain switching and the Ising magnetization.
The obtained piezomagnetic coefficients are fairly large, implying that Ising ferromagnets are promising frontiers when seeking for materials with large piezomagnetic responses.
\end{abstract}

\maketitle


Cross correlations, coupling between physical quantities with orthogonal symmetry properties in systems lacking the corresponding symmetry, have been attracting much attention~\cite{Spaldin2005_Hecman_diagram}.  
In particular, breakings of the most fundamental symmetries, namely the inversion and time-reversal symmetries, have been studied extensively, and novel cross-correlation phenomena such as magneto-electric effects in multiferroics have been established~\cite{Tokura2010.AdvMater.22.1554, Spaldin2019.NatureMater.18.203}. 
Nevertheless, there are many unexplored cross correlations, which are worth extensive investigations both for fundamental and applicational points of view.

One of such unexplored cross correlations is the piezomagnetism (PZM). 
The PZM or piezomagnetic effect refers to the phenomenon that the strain $\varepsilon$ of a magnetic material  responds linearly to the external magnetic field $H$ (i.e. $\varepsilon \propto H$), or its inverse effect, namely the magnetization $M$ induced linearly by external stress $\sigma$ ($M \propto \sigma$). 
The former is also called the linear magnetostriction. 
These effects were first predicted in 1928~\cite{Voigt1928} and its basic theory was established in 1956~\cite{TAVGER1956_1954submitsion}.
It is now understood that materials having symmetry groups without independent time-reversal operation or those with time-reversal operation but only in combination with lattice reflections or rotations can exhibit PZM~\cite{Borovikromanov1994}. 
Experimentally, PZM was first discovered in antiferromagnets $\mathrm{CoF_2}$ and $\mathrm{MnF_2}$ in 1959~\cite{Borovik1959.SovPhysJETP.9.1360,borovik1960piezomagnetism} following theoretical prediction~\cite{Dzyaloshinskii1958.SovPhysJETP.6.621}. 
In these materials, time-reversal symmetry is lost due to the characteristic magnetic structure with up and down magnetic moments sitting respectively on different crystalline sublattices.
This is in clear contrast with ordinary antiferromagnets, whose symmetry groups possess time-reversal operation coupled with lattice translations.
Recently, non-collinear antiferromagnet $\mathrm{UO_2}$ is reported to exhibit hard PZM with a coercive field of as large as 18~T~\cite{jaime2017piezomagnetism}.
More recentry, PZM is attracting renewed attention as a powerful tool to detect non-trivial time-reversal-symmetry breaking (TRSB) in novel states such as magnetic multipole orders~\cite{Meng2023.arXiv2301.06401_Mn3Sn} and altermagnetism~\cite{Aoyama2023.arXiv2305.14786,McClarty2023.arXiv2308.04484}.
Nevertheless, PZM has been only reported in several limited materials so far~\cite{Borovikromanov1994,jaime2017piezomagnetism,Meng2023.arXiv2301.06401_Mn3Sn,Aoyama2023.arXiv2305.14786}, and urges further investigations. 

As mentioned above, PZM is allowed in systems that exhibit TRSB~\cite{TAVGER1956_1954submitsion,Borovikromanov1994}. 
Because of this principle, we can easily predict that ferromagnets should also exhibit piezomagnetism. 
However, surprisingly, PZM has been observed only in antiferromagnets~\cite{jaime2017piezomagnetism,Meng2023.arXiv2301.06401_Mn3Sn,Aoyama2023.arXiv2305.14786,McClarty2023.arXiv2308.04484}.
Observation of PZM in ferromagnets is perhaps hindered by the complex domain configurations with various magnetization directions. 
In such multi-domain states, bulk PZM would be cancelled among domains with opposite magnetic moments.
In addition, ordinary ferromagnetic (FM) magnetostriction, namely strain due to domain configuration change, dominates~\cite{Cullity1971}. 
Moreover, if directions of magnetic moments vary due to domain formation and/or applied magnetic fields, the magnetic point group can also change, making detection of PZM even more complicated. 
Thus, most ferromagnets exhibit ordinary magnetostriction behavior approximated as $\varepsilon \propto (H-H_{\mathrm{coer}})^2$ in magnetic fields near the coercive field $H_{\mathrm{coer}}$~\cite{BrasText}. 
Therefore, it is not straightforward to detect the naively expected piezomagnetism in ferromagnets. 

In this Letter, we report magnetostriction in the itinerant ferromagnet URhGe measured with the multi-axis simultaneous strain measurement technique based on fiber Bragg grating (FBG). 
We found unusual ``V-shaped'' magnetostriction in the FM phase only for specific combinations of field and strain directions. 
This is attributed to the ferromagnet PZM, which has been overlooked for more than a half century.

Our target material is the Ising-like itinerant ferromagnet URhGe. 
URhGe has the orthorhombic $Pnma$ crystal structure~\cite{Tran1998.JMagMagMater186.81}. 
This material exhibits ferromagnetism below the Curie temperature $\TC = 9.5$~K~\cite{DeBoer1990.PhysicaB.163.175,Tran1998.JMagMagMater186.81} and subsequently superconductivity below $\Tsc = 0.28$~K~\cite{Aoki2001.Nature.413.613}. 
Due to the strong spin-orbit coupling, the magnetic moment in URhGe shows strong Ising features with the easy axis along the $c$ direction~\cite{Aoki2019_Review}. 
Such Ising nature is inherited in the FM state. 
Recently, Mineev discussed PZM in the FM and superconducting states of URhGe and its related materials UCoGe and UGe$_2$~\cite{Mineev2021}. 
Because of the Ising magnetic antisotropy, the magnetic point group of the FM state is well-defined to be $D_{2h} (C_{2h})$ irrespective of magnetic-field directions as long as the field is not too strong.
For this magnetic point group, PZM is indeed allowed; the piezomagnetic effect obeys
$\varepsilon_{\mu} = \sum_k Q_{k\mu}H_k$ with the non-vanishing piezomagnetic tensor 
\begin{align}
    Q_{k\mu}
    =
    \begin{pmatrix}
        0 & 0 & 0 & 0 & Q_{15} & 0 \\
        0 & 0 & 0 & Q_{24} & 0 & 0 \\
        Q_{31} & Q_{32} & Q_{33} & 0 & 0 & 0 \\
    \end{pmatrix},
\label{eq:PZM-tensor}
\end{align}
where $\varepsilon_\mu$ is the strain expressed using the Voigt notation  ($\varepsilon_1$ corresponding $\varepsilon_{aa}$, etc), and $H_k$ is the $k$-component of the magnetic field ($k = 1$, $2$, and $3$ corresponding to the $a$, $b$, and $c$ directions)~\cite{Borovikromanov1994,Mineev2021}.
This tensor indicates that $c$-axis field will produce piezomagnetic linear magnetostriction in $\varepsilon_{aa}$, $\varepsilon_{bb}$, and $\varepsilon_{cc}$,
whereas no PZM in these normal strains for fields along $a$ or $b$ axis.


For the present study, we used a high-quality single crystals of URhGe prepared with the Czochralski method.
Strain $\varepsilon_{bb}$ and $\varepsilon_{cc}$ of this sample was simultaneously measured by using FBG sensors.
The FBG is a periodic grating embedded to the core of an optical fiber and we can sensitively measure the strain from the change in the wavelength of the light reflected from an FBG pasted to the sample~\cite{Jaime2017_FBG}. 
The strain transmission rate between the sample and FBG through glue is calibrated based on thermal-expansion measurements. 
To introduce light to FBGs and measure the spectra of reflected light, we used a commercial interrogator (KYOWA EFOX-1000B-4). 
The strain measurements were performed in a commercial cryostat (Quantum Design, PPMS), whereas magnetization $M$ was measured with a commercial magnetometer (Quantum Design, MPMS). 
Details of experimental method are explained in Supplemental Material~\cite{SuppleMat}.

\begin{figure}[t]
\begin{center}
\includegraphics[width=8.5cm]{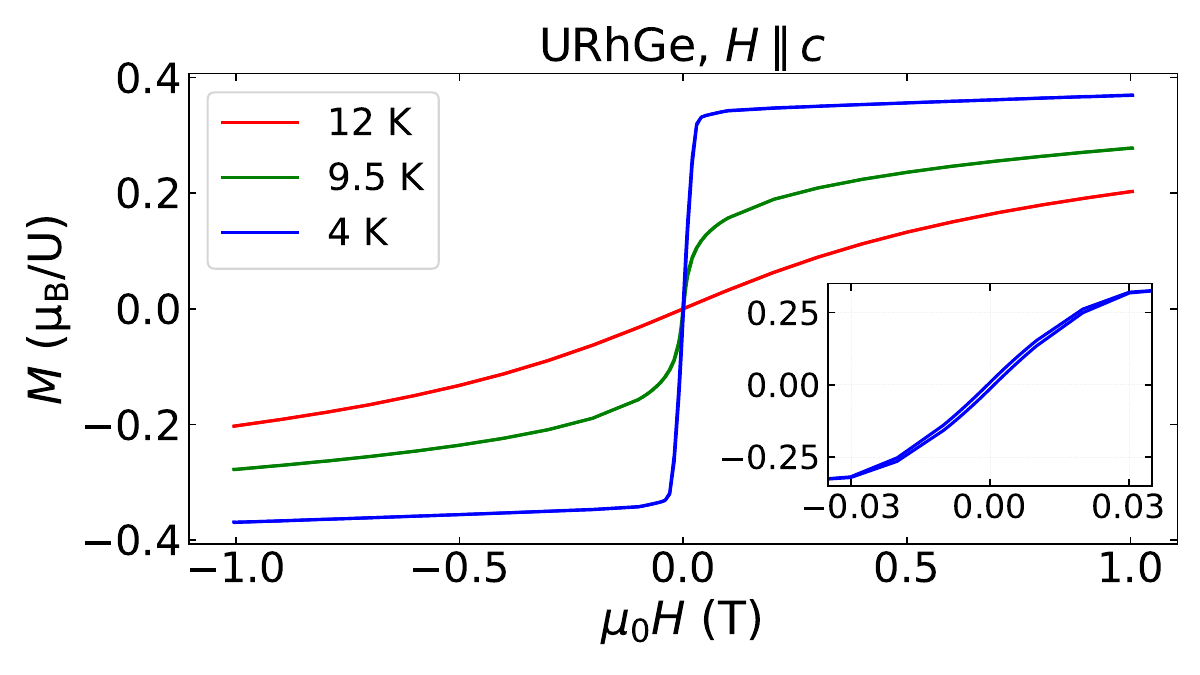}
\caption{(color online) Magnetization $M$ of URhGe measured with $H \parallel c$. 
The inset is an enlarged view of the 4.0-K data around $H = 0$. 
Below $\TC$, $M(H)$ exhibits step-function like behavior with a narrow hysteresis, indicating that the multi-domain state is almost negligible.
\label{fig:magnetization}}
\end{center}
\end{figure}

First, we show in Fig.~\ref{fig:magnetization} $M$ of the URhGe sample for $H$ along the easy-magnetization axis ($c$ axis) measured at various temperatures. 
In the FM state below \tc $\sim$ 9.5~K, $M(H)$ exhibits step-like change around $H=0$ with very narrow hysteresis width.
The step-like change indicates that all magnetic moments flip simultaneously without forming FM domains.
Thus, this compound is free from complicated phenomena originating from multi-domain configurations.
Note that the saturated moment is about 0.4~$\mu_{\mathrm B}$/U, which is much smaller than the value expected for U $5f^2$ or $5f^3$ configuration ($\sim 3~\mu_{\mathrm B}$/U), indicating the weak  itinerant ferromagnetism.

\begin{figure}[t]
\begin{center}
\includegraphics[width=8.5cm]{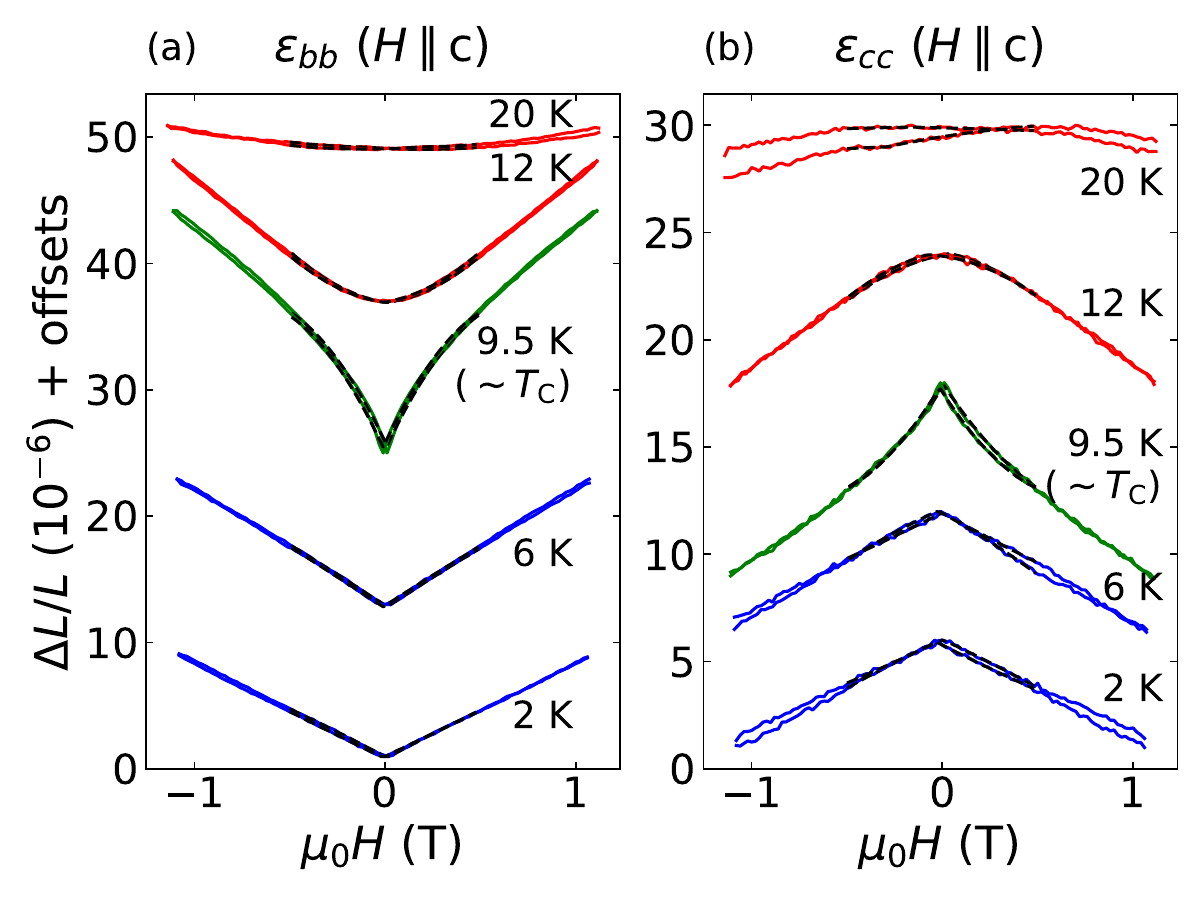}
\caption{(color online) Magnetostriction of URhGe in (a) $\varepsilon_{bb}$ and (b) $\varepsilon_{cc}$ measured under $H\parallel c$. 
Each curve is vertically offset for clarity.
Below $\TC$ (blue curves), V-shaped magnetostriction originating from PZM is observed in both strain components, in clear contrast to the quadratic behavior above $\TC$ (red curves) or the nonlinear behavior with a cusp due to critical behavior at $\TC$ (green curves).
Results of fittings using Eq.~\eqref{eq:fitting} are shown with black dotted curves.
\label{fig:Hc}}
\end{center}
\end{figure}

Next, in Fig.~\ref{fig:Hc}, we show our main result, namely the magnetostriction measured under $H \parallel c$ at various temperatures. 
Above $\TC$ (red curves), both $\varepsilon_{bb}$ and $\varepsilon_{cc}$ exhibit quadratic behavior around zero field.
This is ordinary behavior for paramagnets; if $M$ is proportional to $H$, the ordinary magnetostriction should obey the relation $\varepsilon\propto M^2 \propto H^2$.
The difference in the sign of the magnetostriction reflects the anisotropic thermal expansion at zero field (Fig.~\ref{fig:thermal}):
When the magnetic moment grows as temperature decreases, $\varepsilon_{bb}$ tends to expand and $\varepsilon_{cc}$ prefers to shrink.
The application of magnetic field, bringing the sample closer to the FM state, causes the same effect.
This tendency that a shorter $b$ axis tends to disfavor the FM order is consistent with the fact that $\TC$ decreases under uniaxial pressure along the $b$ axis~\cite{Braithwaite2018.PhysRevLett.120.037001}.
This uniaxial-pressure effect is known to be consistent with the negative jump in the $b$-axis thermal-expansion coefficient at $\TC$~\cite{AOKI2011573_strain} through the Ehrenfest relation.

At $\TC = 9.5$~K (green curves), $\varepsilon_{bb}$ and $\varepsilon_{cc}$ show sharp kinks at $H=0$, followed by non-zero curvatures at finite fields.
This at first glance looks anomalous but is mostly attributable to the ordinary magnetostriction $\varepsilon\propto M^2$ together with the critical behavior in $M$ at $\TC$:
Mean-field theories predict $M\propto H^{1/3}$ at the transition temperature~\cite{BlundellText}, hence $\varepsilon$ is expected to show nonlinear $H^{2/3}$ behavior.
Indeed, such cusp-like magnetostriction at $\TC$ has been reported in other ferromagnets such as UIr~\cite{Knafo2009.JPhysSocJpn.78.043707} and TbCo$_2$Mn$_x$~\cite{Gerasimov2021_ferro_example}.
The behavior at $\TC$ will be discussed again later.

Interestingly, in the FM state (blue curves), $\varepsilon_{bb}(H)$ and $\varepsilon_{cc}(H)$ curves are strikingly V-shaped, with robust $H$-linear dependences in both $H>0$ and $H<0$ regions.
As a consequence, $\varepsilon$ is proportional to $|H|$.
This behavior remains down to 2~K, the lowest temperature of the present study.
Compared with the step-like $M(H)$ curve in the FM phase (Fig.~\ref{fig:magnetization}), it is clear that the $\varepsilon\propto |H|$ dependence cannot be described by the ordinary magnetostriction $\varepsilon\propto M^2$, as demonstrated in Figs.~\ref{fig:URhGe_M2_and_strain_b} and \ref{fig:URhGe_M2_and_strain_c}~\cite{SuppleMat}.
Below, we will discuss other features to conclude that this V-shaped magnetostriction originates from the PZM in URhGe.

\begin{figure}[t]
\begin{center}
\includegraphics[width=8.5cm]{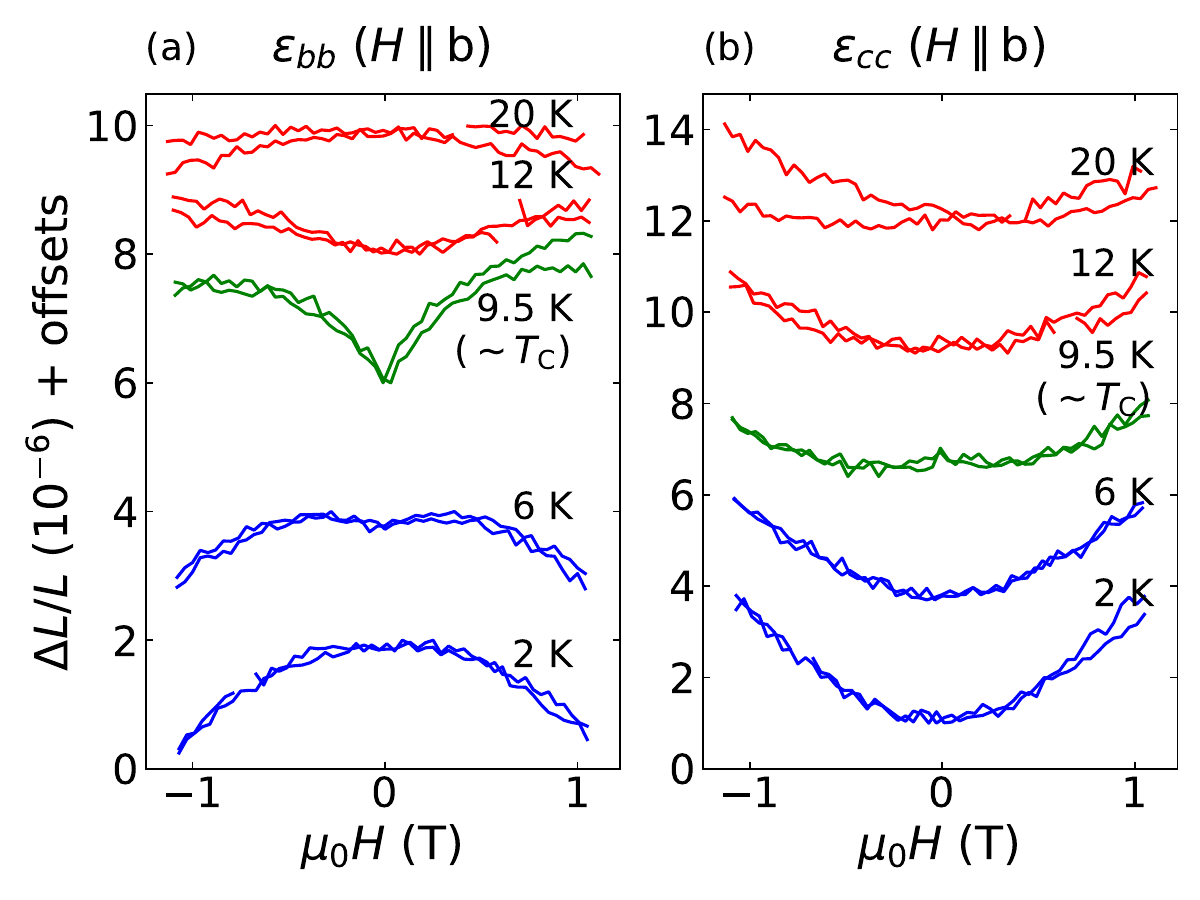}
\caption{(color online) Magnetostriction of URhGe in (a) $\varepsilon_{bb}$ and (b) $\varepsilon_{cc}$ measured under $H\parallel b$. 
Each curve is vertically offset for clarity.
V-shaped magnetostriction was not observed below $\TC$ (blue curves), whereas the behavior above $\TC$ (red curves) and at $\TC$ (green curves) qualitatively resembles those observed under $H\parallel c$ (Fig.~\ref{fig:Hc}).
\label{fig:Hb}}
\end{center}
\end{figure}

We firstly examine the anisotropy of the V-shaped magnetostriction.
We show in Fig.~\ref{fig:Hb} the magnetostriction measured under $H \parallel b$.
The strain above $\TC$ is weak and quadratic against the magnetic fields. 
At $T=\TC$, the strain showed a cusp at $H=0$, which is similar to those observed in $H\parallel c$ and is attributable to the critical behavior.
For $T<\TC$, the $\varepsilon(H)$ curves become quadratic again, in clear contrast to the V-shaped curves observed in $H\parallel c$.
This quadratic behavior is consistent with the previous report of $\varepsilon_{bb}$ measured under $H\parallel b$~\cite{Nakamura2018.ProgNuclSciTechnol.5.123}.
Thus, the V-shaped magnetostriction has strong anisotropy depending on the field direction.
This anisotropy is consistent with the PZM tensor shown in Eq.~\eqref{eq:PZM-tensor}:
for URhGe, $Q_{21}$, $Q_{22}$, and $Q_{23}$ are zero, meaning that no PZM should occur in $\varepsilon_{aa}$, $\varepsilon_{bb}$, or $\varepsilon_{cc}$ under $H\parallel b$.

\begin{figure}[t]
\begin{center}
\includegraphics[width=8.5cm]{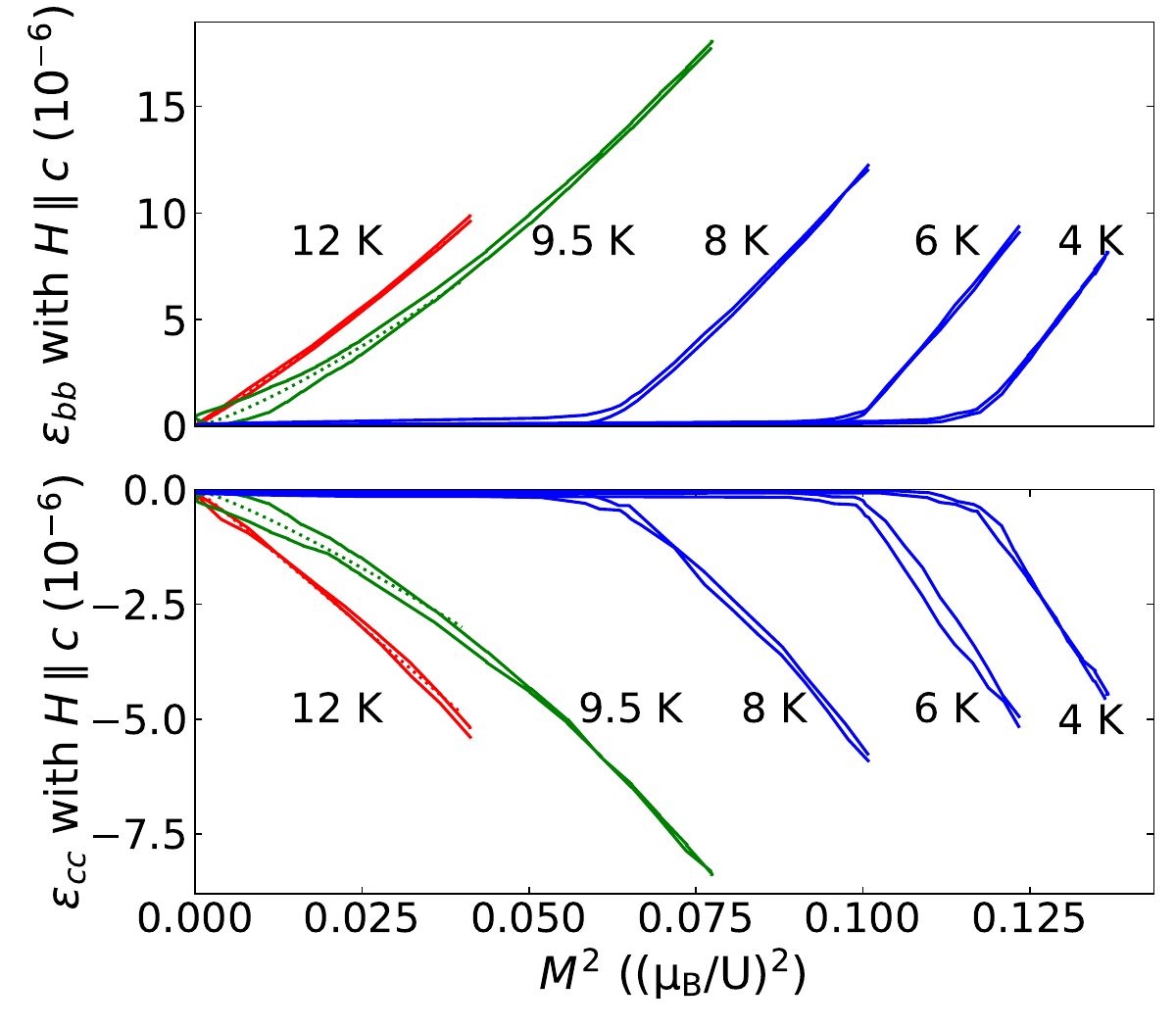}
\caption{(color online) Relation between $\varepsilon_{cc}$ and $M^2$ under $H\parallel c$ at various temperatures deduced from the data in Figs.~\ref{fig:magnetization}-\ref{fig:Hb}.  
The strain is almost proportional to $M^2$ above $\TC$ (red solid curve) as demonstrated by the linear fit shown with the red dotted line.
At $\TC$ (green curve), the curvature is non-zero, but the power-law relation still holds between strain and magnetization.
In clear contrast, there is no simple power-law relation between them in the FM phase (blue curves).
\label{fig:strain_VS_M2}}
\end{center}
\end{figure}

Secondly, we compare $\varepsilon$ and $M$ in more detail.
For ordinary magnetostriction, the empirical relation $\varepsilon\propto M^2$ often holds.
We thus plot $\varepsilon_{bb}$ and $\varepsilon_{cc}$ as functions of $M^2$ in Fig.~\ref{fig:strain_VS_M2}.
One can clearly see that, above $\TC$, the strain is nearly proportional to $M^2$.
At $\TC$, where the cusps in $\varepsilon(H)$ curves at $H=0$ are observed, the strain vs $M^2$ curves acquire non-zero curvature, but maintain smooth relations close to linear.
Indeed, when we fit the curves with $\varepsilon\propto M^{2\alpha}$ using the exponent $\alpha$ as the fitting parameter, we obtain $\alpha = 1.095$ for $T = 12$~K and $1.268$ for $T = 9.5$~K, both being close to unity.
This result manifests that the strain in URhGe above and at $\TC$ is attributed mainly to the ordinary magnetostriction.
In contrast, the $\varepsilon(M^2)$ curve at 4.0~K does not show power-law relation.
Thus, the V-shaped magnetostriction does not have the conventional relation to the magnetization.

%

Now we explain that the observed V-shaped magnetostriction is indeed well attributable to PZM.
Generally in magnets showing TRSB, two magnetic structures connected by the time-reversal operation exhibit opposite signs of piezomagnetic coefficients, respectively.
Thus, in typical piezomagnetic materials having the piezomagnetic coefficient $Q$ for one of the magnetic structures, the strain obeys $\varepsilon = QH$ as long as the magnetic structure is kept, while the opposite behavior $\varepsilon = -QH$ emerges when the magnetic structure is reversed by strong magnetic field exceeding $H_{\mathrm{coer}}$.
This results in butterfly-like $\varepsilon(H)$ curves~\cite{jaime2017piezomagnetism,Meng2023.arXiv2301.06401_Mn3Sn}.
In URhGe under $H\parallel c$, $H_{\mathrm{coer}}$ is nearly zero, as demonstrated in the step-like $M(H)$ curve (Fig.~\ref{fig:magnetization}).
Zero coercive field changes the butterfly-like $\varepsilon(H)$ curve to the V-shaped curve, as observed in URhGe.

These analyses and considerations confirm that the V-shaped magnetostriction observed under $H\parallel c$ in URhGe originates from PZM, marking the first clear example of PZM in ferromagnets.
It is interesting that, although ferromagnets are most archetypal examples exhibiting spontaneous TRSB, required feature to realize PZM as theoretically established in 1956~\cite{TAVGER1956_1954submitsion}, ferromagnetic PZM has been overlooked for more than 60 years.
The near absence of multi-domain state near $H=0$ and the strong Ising nature fixing the magnetic point group irrespective of the field directions are the keys to avoid various complications characteristic to ferromagnets.

\begin{figure}[t]
\begin{center}
\includegraphics[width=8.5cm]{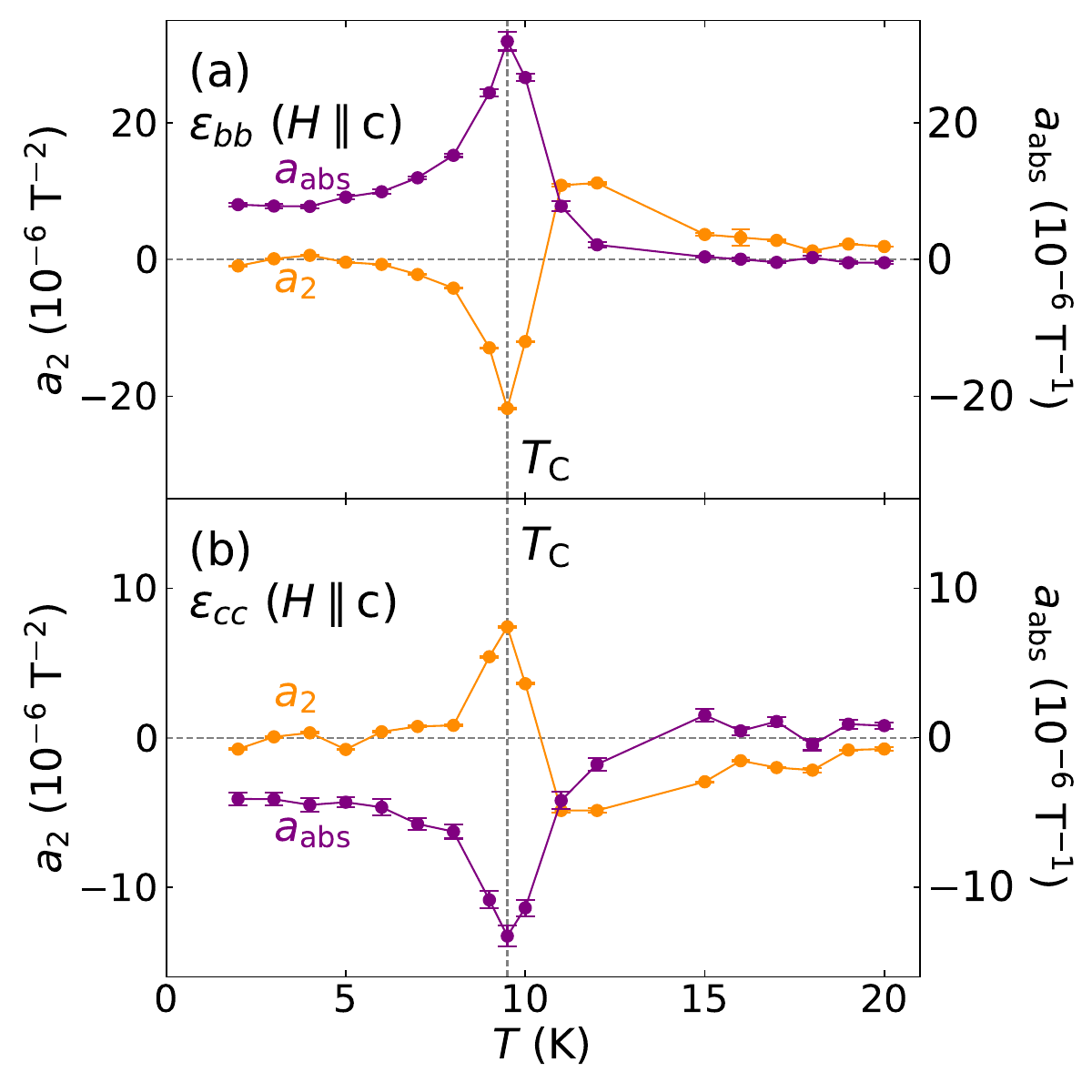}
\caption{(color online) Temperature dependence of the fitting coefficients $a_2$ and $a_\mathrm{abs}$ (Eq.~\eqref{eq:fitting}) for the magnetostrictions (a) $\varepsilon_{bb}$ and (b) $\varepsilon_{cc}$ measured under $H \parallel c$. 
The vertical dotted line indicate $\TC$.
The dominant magnetic field dependence of the strain changes from $\varepsilon\sim H^2$ (ordinary magnetostriction) above $\TC$ to $\varepsilon \sim |H|$ (PZM) below $\TC$.
The singular behaviors at $\TC$ is attributable to the critical non-linear behavior in $\varepsilon(H)$ as discussed in the text.
\label{fig:fitting}}
\end{center}
\end{figure}

To investigate the temperature evolution of the PZM, we fitted the data with the function
\begin{align}
    \varepsilon(H) = \varepsilon_0 + a_2 (\mu_0 H)^2 + a_\mathrm{abs}|\mu_0H|
\label{eq:fitting}
\end{align}
in the field range $-0.5~\mathrm{T}\leq \mu_0H\leq +0.5~\mathrm{T}$.
As shown with the dotted curves in Fig.~\ref{fig:Hc}, the fittings are successful for all data sets.
The temperature dependence of $a_2$ and $a_\mathrm{abs}$ are shown in Fig~\ref{fig:fitting}. 
Above $\TC$, $a_2$ is dominant, whereas $a_{\mathrm{abs}}$ is within the noise level. 
The peaks at $\TC$ are attributable to the critical behavior in $\varepsilon(H)$ discussed in previous paragraphs: when we are forced to fit the cusp-like $\varepsilon(H)$ curve using Eq.~\eqref{eq:fitting}, we mathmatically need a large $a_{\mathrm{abs}}|H|$ term and similarly large $a_{2}H^2$ term with the opposite sign.
Much below $\TC$, $a_\mathrm{abs}$ becomes dominant and $a_2$ is nearly zero, as expected from the V-shaped magnetostriction.
These results quantitatively show that the behavior of strain drastically changes from the ordinary magnetostriction $\varepsilon\propto H^2$ above $\TC$ to the piezomagnetic response $\varepsilon\propto |H|$ in the FM phase.

The fitting coefficient $a_\mathrm{abs}$ in the FM phase is equivalent to the corresponding piezomagnetic-tensor component. 
Taking the lowest-temperature values in the FM state, we obtain 
$Q_{32}=8.0 \times 10^{-6}~\mathrm{T}^{-1}$ and $Q_{33}=-4.1 \times 10^{-6}~\mathrm{T}^{-1}$ for URhGe. 
These values are compared with results of other piezomagnets. 
As listed in Table~\ref{tab:other_piezomagnets}, most of the known piezomagnets exhibit $|Q_{k\mu }|$ of less than $2 \times 10^{-6}~\mathrm{T}^{-1}$, whereas recently found piezomagnets such as $\mathrm{UO_2}$ and $\mathrm{Mn_3Sn}$ exhibit $|Q_{k\mu}|$ exceeding $10\times 10^{-6}~\mathrm{T}^{-1}$~\cite{jaime2017piezomagnetism,Meng2023.arXiv2301.06401_Mn3Sn}.
The observed PZM components of URhGe reach around 50\% of these values.
This comparison implies that ferromagnets with appropriate conditions are candidate materials realizing large piezomagnetic coefficients.
We comment that the signs of $Q$ is well-defined in our case due to the very soft magnetism, whereas other piezomagnets exhibit PZM of both signs depending on magnetic structures.

\begin{table}[h]
\caption{Properties of known piezomagnetic materials~\cite{Borovikromanov1994, jaime2017piezomagnetism, Aoyama2023.arXiv2305.14786, Meng2023.arXiv2301.06401_Mn3Sn}. 
The values with $^\ast$ were obtained by using strain-induced magnetization measurements, while the others by linear-magnetostriction measurements.
In the column of magnetization, AFM refers to an antiferromagnet, MOO a magnetic-octupole ordered material, ALM an altermagnet, and FM a ferromagnet.
\label{tab:other_piezomagnets}}
\begin{center}
\begin{ruledtabular}
\begin{tabular}{ccccc}
Materials & $Q_{k\mu}$ ($10^{-6}~\mathrm{T}^{-1}$) & $T$ (K) & Magnetism & Refs. \\
\hline
CoF$_2$               & $Q_{14}= \pm 21^\ast$    & 20 & AFM & \cite{borovik1960piezomagnetism}\\
                      & $Q_{36} = \pm 9.8$       & 4  &     &\cite{Prokhorov1969_CoF2_LM}\\
 MnF$_2$              & $Q_{14} = \pm 0.2^\ast$  & 20 & AFM &\cite{borovik1960piezomagnetism}\\
                      & $Q_{14} = \pm 0.07^\ast$ & 60 &     &\cite{Baruchel1988_MnF2}\\
 $\alpha$-Fe$_2$O$_3$ & $Q_{22} = \pm 1.9$       & 78 & AFM &\cite{ANDRATSKil1966_Fe2O3}\\
                      & $Q_{22} = \pm 3.2^\ast$  & 77 &     &\cite{Anderson1964_Fe2O3}\\
                      & $Q_{22} = \pm 1.3$       & 100&     &\cite{Levitin1973_Fe2O3}\\
                      & $Q_{23} = \pm 2.5^\ast$  & 292& canted AFM&\cite{ANDRATSKil1966_Fe2O3}\\
 YFeO$_3$             & $Q_{15} = \pm 1.7$       & 6  & AFM &\cite{Kadomtseva1981_YFeO3}\\
 YCrO$_3$             & $Q_{15} \simeq \pm 1$    & 6  & AFM &\cite{Kadomtseva1981_YCrO3}\\
 DyFeO$_3$            & $Q_{36} = \pm 6.0$       & 6  & AFM &\cite{Zvezdin1985_DyFeO3}\\
 UO$_2$               & $Q_{14} = \pm 10.5$      & 2.5& AFM &\cite{jaime2017piezomagnetism}\\
 MnTe                 & $Q_{\mathrm{ave}}=0.0050^\ast$ & 250& AFM / ALM&\cite{Aoyama2023.arXiv2305.14786}\\
 Mn$_3$Sn             & $Q_{11} = \pm 4.4^\ast$  & 300& AFM / MOO &\cite{Ikhlas2022_Mn3Sn}\\
                      & $Q_{11} = \pm 14.6$      &    &           &\cite{Meng2023.arXiv2301.06401_Mn3Sn}\\
URhGe                 & $Q_{32} = +8.0$          &  2 & FM        & This work\\
                      & $Q_{33} = -4.1$          &    &           &  \\
\end{tabular}
\end{ruledtabular}
\end{center}
\end{table}

One open question is the microscopic mechanism of the observed large PZM in URhGe. 
Naively, sufficient magneto-lattice coupling is necessary to induce PZM.
Such magneto-lattice coupling should originate from spin magnetic moments with strong spin-orbit coupling (SOC) and/or orbital magnetic moments. 
Uranium compounds, exhibiting strong SOC and thus angular-momentum coupling between spin and orbitals, should satisfy both criteria.
The Ising magnetic nature in URhGe indeed originates from SOC~\cite{Wilhelm2017.PhysRevB.95.235147}.
More recently, it is revealed that the non-symmorphic crystalline structure, resulting from the zig-zag uranium chain, leads to anisotropic pseudo-spin texture pointing perpendicularly to the Brillouin-zone boundaries under SOC~\cite{Suh2023.PhysRevResearch.5.033204}.
Such pseudo-spin texture may explain microscopic origins of the Ising ferromagnetism in URhGe, and may further provide bases toward clarifying mechanism of PZM.

To summarize, we revealed the piezomagnetism (PZM) in the itinerant Ising ferromagnet URhGe. 
To our knowledge, this is the first report of ferromagnetic PZM, which has been overlooked for many years.
The observed piezomagnetic coefficients range around 50\% of the largest ones ever known.
This work demonstrates that Ising ferromagnets without multi-domain states can be good candidates when seeking for materials with large piezomagnetic responses, which can be utilized for novel sensors or actuators.
This new finding would stimulate further experimental and theoretical studies toward understanding of microscopic mechanisms of PZM, which can be a novel probe of TRSB phenomena occurring in various materials.

\begin{acknowledgments}
We acknowledge V.~P.~Mineev, Y.~Yanase, G.~Mattoni, S.~Kitagawa, D.~Agterberg, K.~Ogushi and Y.~Maeno for fruitful discussion.
The work at Kyoto Univ.\ was supported by Grant-in-Aids for 
Scientific Research on Innovative Areas ``Quantum Liquid Crystals'' (KAKENHI Grant Nos.~JP20H05158, JP22H04473) from the Japan Society for the Promotion of Science (JSPS),
Academic Transformation Area Research (A) ``1000 Tesla Science’’ (KAKENHI Grant No.~23H04861),
Grant-in-Aids for Scientific Research (KAKENHI Grant Nos.~JP17H06136, JP20H00130, JP22H01168) from JSPS,
by Core-to-Core Program (No.~JPJSCCA20170002) from JSPS,
by Bilateral Joint Research Projects (No.~JPJSBP120223205) from JSPS,
by research support fundings from The Kyoto University Foundation, 
by ISHIZUE 2020 and 2023 of Kyoto University Research Development Program, and by Murata Science and Education Foundation.
The work at Tohoku Univ.\ was supported by Grant-in-Aids for Scientific Research (KAKENHI JP22H04933, JP20KK0061, and JP20K20889).
S.~Yonezawa acknowledges support for the construction of experimental setups from Research Equipment Development Support Room of the Graduate School of Science, Kyoto University; and support for liquid helium supply from Low Temperature and Materials Sciences Division, Agency for Health, Safety and Environment, Kyoto University.
\end{acknowledgments}



%



\clearpage

\renewcommand{\theequation}{S\arabic{equation}}
\setcounter{equation}{0}
\renewcommand{\thefigure}{S\arabic{figure}}
\setcounter{figure}{0}
\renewcommand{\thetable}{S\arabic{table}}
\setcounter{table}{0}
\renewcommand{\thepage}{S\arabic{page}}
\setcounter{page}{1}
\renewcommand{\thesubsection}{S\arabic{subsection}}
\setcounter{subsection}{0}
\onecolumngrid

\section*{Supplemental Material}
\subsection{Details of experimental methods} 

In this study, we used a high-quality single crystal of URhGe grown with the Czochralski method in a tetra-arc furnace. 
A single crystal ingot was oriented by taking Laue photographs, and then cut to the dimensions $1.49 \times 1.46 \times 1.48~\mathrm{mm}^3$ using a spark cutter. 
The crystal was subsequently annealed under ultra high vacuum at 900${}^\circ$C for 10 days.
Before strain measurements, we investigated field and temperature dependence of the magnetization using a commercial magnetometer (Quantum Design, MPMS) in order to characterize magnetic properties of the sample.

We used FBG as an optical-fiber-based strain sensor. 
FBG is a periodic grating embedded to the core an optical fiber.
Small modulation of the refractive index $n$ is formed with the period $d$.
When light is introduced to this core with the grating, only light with the particular Bragg wavelength ($\lambda_\mathrm{B} = 2nd$) is reflected while others transmit through the fiber. 
To measure the strain of a sample, FBG is pasted to the sample using a glue. 
The Bragg wavelength shifts due to changes in $d$, which follows the expansion or shrinkage of the sample.  
Thus we can detect the strain of the sample from the shift of the reflected Bragg wavelength.  
Including additional contribution due to small changes in $n$ caused by strain and temperature, we obtain the relation between strain and wavelength as
\begin{align}
    \frac{\Delta L}{L}
    =
    \frac{K}{0.78}
    \left(
    \frac{\Delta \lambda_\mathrm{B}}{\lambda_\mathrm{B}}
    -
    \frac{\Delta \lambda_\mathrm{B,free}}{\lambda_\mathrm{B,free}}
    \right),
\end{align}
where $\lambda_\mathrm{B,free}$ is the Bragg wavelength of the FBG not pasted to the sample, $K^{-1}$ is the strain transmission rate that will be explained later, and the factor 0.78 originates from strain-optic tensor~\cite{Jaime2017_FBG}.

\begin{figure}[b]
\begin{center}
\includegraphics[width=8cm]{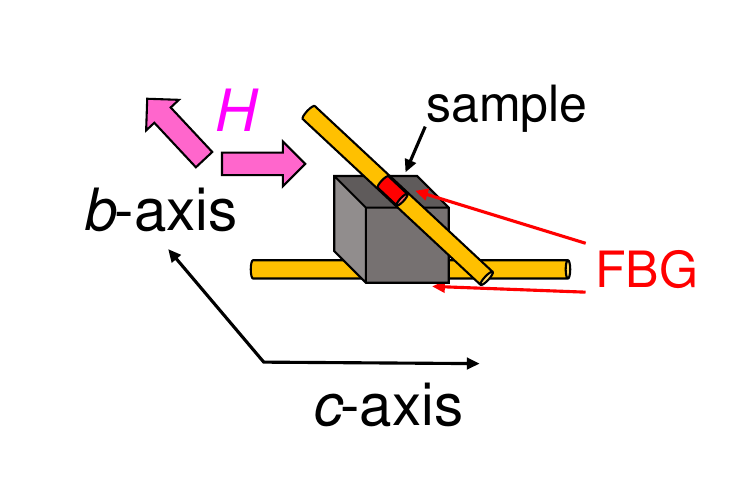}
\caption{Schematic to explain how to paste FBG sensors to the sample. 
We pasted two FBGs parallel to the $b$ and $c$ axes of the sample, respectively, in order to perform multi-axis simultaneous strain measurements.
We performed two sets of experiments with the applied magnetic field along either the $b$ or $c$ axis.
\label{fig:sample}}
\end{center}
\end{figure}

In this study, we used FBGs made of a bending insensitive fiber. 
One fiber contains four FBGs whose Bragg wavelengths were 1530~nm, 1540~nm, 1550~nm, and 1560~nm. 
Two of the FBGs were pasted along the $b$ and $c$ axes of the sample with cyanoacrylate glue (Konishi \#31701) as illustrated in Fig.~\ref{fig:sample}.  
The other FBGs were placed in free space close to the sample and the light wavelength reflected from them, $\lambda_\mathrm{B,free}$, were used for the background calibration. 
Only a corner of the sample was fixed to the sample stage by varnish (GE7301) in order to minimize extrinsic strain from the sample stage. 
To measure $\lambda_\mathrm{B}$, we used a commercial interrogator (KYOWA EFOX-1000B-4), which is equipped with the variable-wavelength laser and an optical spectrometer. 
The scanning frequency of the interrogator was set to be 100 Hz. 
The sample was cooled down with a commercial cryostat (Quantum Design, PPMS) equipped with a horizontal superconducting  split-coil magnet. 
To measure the accurate temperature and magnetic field in the sample space, we placed a thermometer (Lake Shore, Cernox CX-1050) and a Hall sensor (Toshiba, THS118) in the vicinity of the sample.
These two sensors were calibrated before measurements.

\begin{figure}[tb]
\begin{center}
\includegraphics[width=10cm]{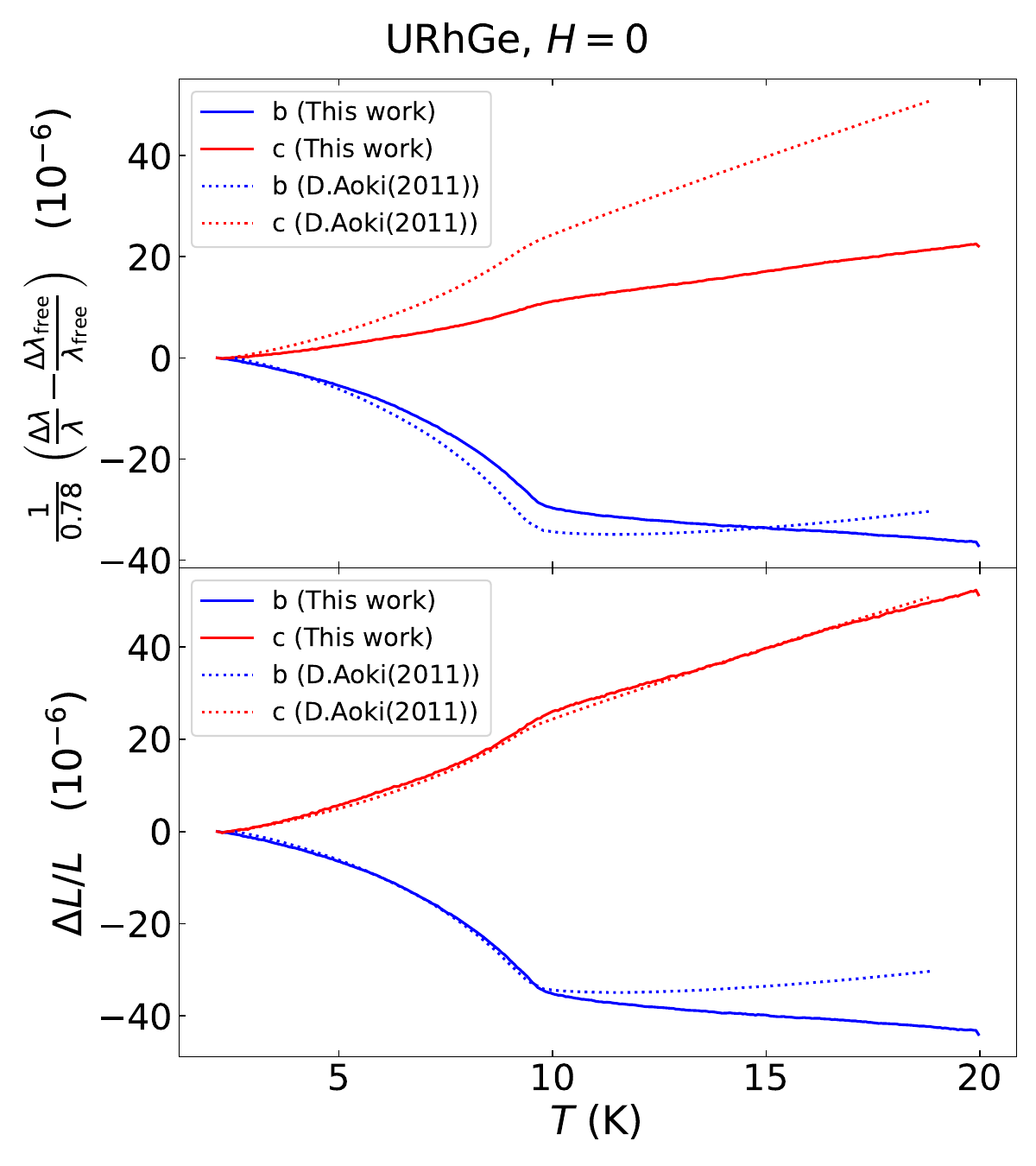}
\caption{(a) Nominal thermal expansion of the URhGe sample measured with FBG (obtained by assuming the strain transmission rate $K^{-1} = 1$), compared with the literature data (Aoki 2011: Ref.~\cite{AOKI2011573_strain}). 
The apparent discrepancy is attributable to the actual strain transmission rate $K^{-1}$ smaller than 1. 
(b) Calibrated thermal expansion data. 
The transmission rate $K^{-1}$ is determined so that the raw data shown in (a) multiplied by $K$ match the literature value. 
From this experiment, we obtained $K_c^{-1} = 0.433$ and $K_b^{-1}=0.84$.
\label{fig:thermal}}
\end{center}
\end{figure}

Before the measurement of magnetostriction, it is necessary to calibrate the strain transmission rate $K^{-1}$. 
For that purpose, we measured zero-field thermal expansion using our setup and compare it with results of a previous research~\cite{AOKI2011573_strain}. 
Figure~\ref{fig:thermal}(a) compares the raw observed strain assuming $K=1$ and the literature data~. 
The apparent difference in the two data sets originates from imperfect transmission of the strain through the glue. 
We determined $K^{-1}$ so that our data multiplied by $K$ matches the literature value. 
As shown in Fig.~\ref{fig:thermal}(b), the corrected data with $K^{-1} = 0.840$ for the $b$ axis and $K^{-1} = 0.433$ for the $c$ axis both match the the literature data in particular below $\TC$.
Strain data shown in the rest of this Letter are corrected by using these $K$ values.

\clearpage

\subsection{Additional experimental data}

\begin{figure}[tbh]
\begin{center}
\includegraphics[width=10cm]{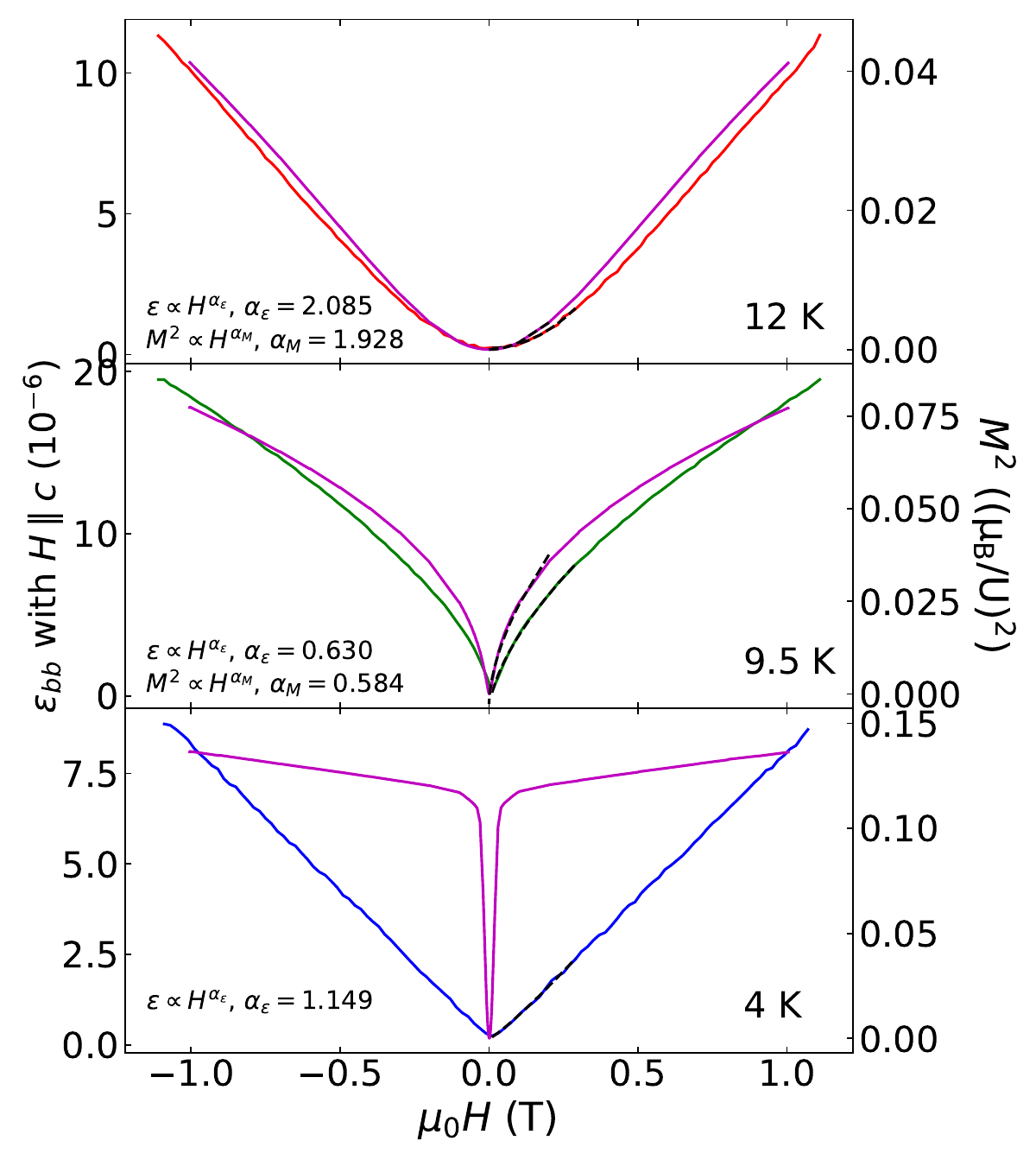}
\caption{Comparison of $\varepsilon_{bb}(H)$ (red, green, blue curves; left axis) and $M^2(H)$ (purple curves; right axis) of URhGe measured under $H\parallel c$ at various temperatures. 
\label{fig:URhGe_M2_and_strain_b}}
\end{center}
\end{figure}

\begin{figure}[tbh]
\begin{center}
\includegraphics[width=10cm]{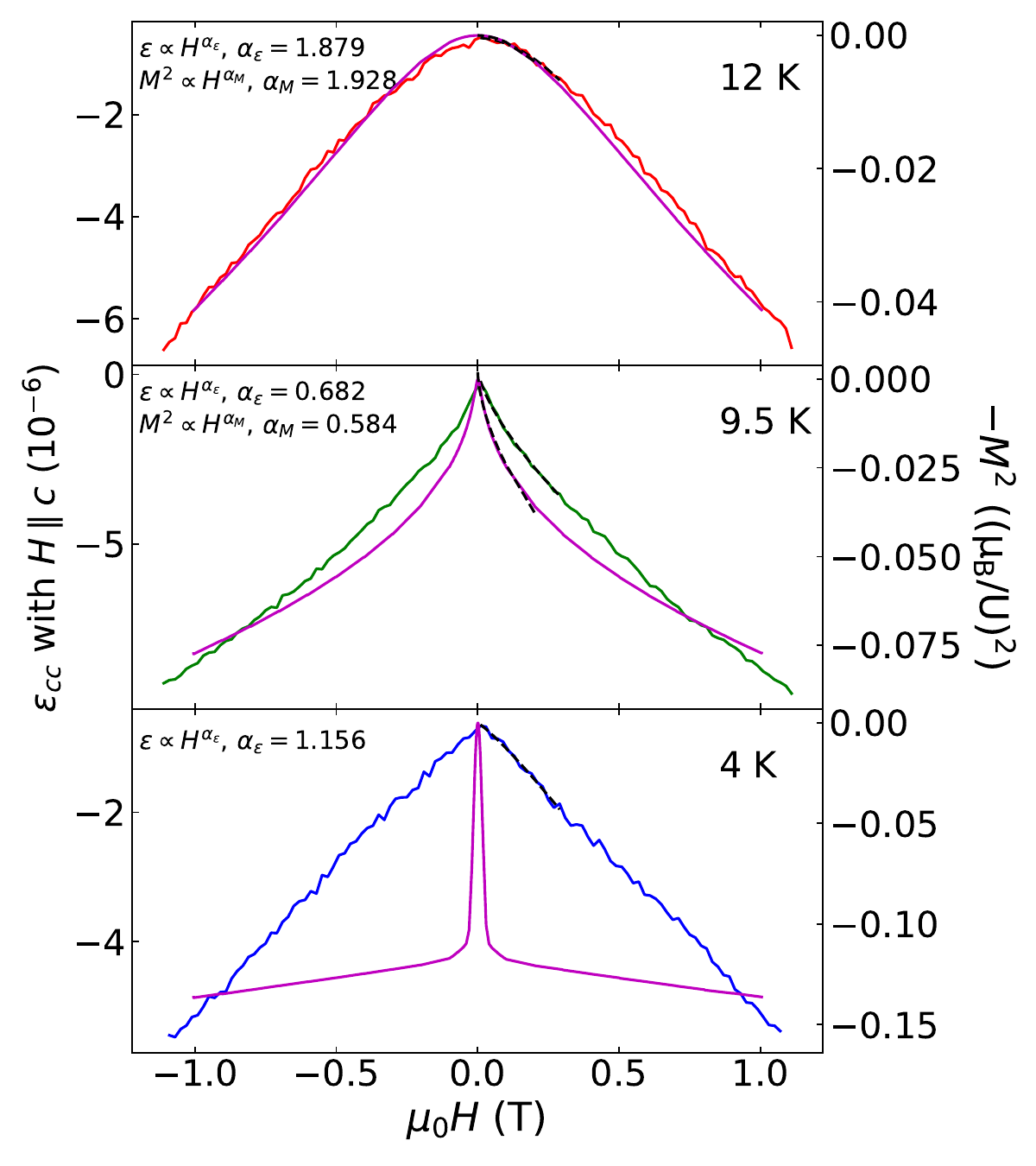}
\caption{Comparison of $\varepsilon_{cc}(H)$ (red, green, blue curves; left axis) and $-M^2(H)$ (purple curves; right axis) of URhGe measured under $H\parallel c$ at various temperatures. 
\label{fig:URhGe_M2_and_strain_c}}
\end{center}
\end{figure}

\end{document}